# Charge-based quantum computing using single donors in semiconductors

L.C.L. Hollenberg*, A. S. Dzurak†, C. Wellard*, A. R. Hamilton†, D. J. Reilly†, G. J. Milburn‡ and R. G. Clark†

*Centre for Quantum Computer Technology*
\* *School of Physics, University of Melbourne, VIC 3010, Australia*
† *Schools of Physics and Electrical Engineering, University of New South Wales, NSW 2052, Australia*
‡ *School of Physics, University of Queensland, QLD 4072, Australia*

Solid-state quantum computer architectures with qubits encoded using single atoms are now feasible given recent advances in atomic doping of semiconductors. Here we present a charge qubit consisting of two dopant atoms in a semiconductor crystal, one of which is singly ionised. Surface electrodes control the qubit and a radio-frequency single electron transistor provides fast readout. The calculated single gate times, of order 50ps or less, are much shorter than the expected decoherence time. We propose universal one- and two-qubit gate operations for this system and discuss prospects for fabrication and scale up.

In the search for an inherently scaleable quantum computer (QC) technology solid-state systems are of great interest. One of the most advanced proposals is based on superconducting qubits[1], where coherent control of qubits has been demonstrated, and decoherence times measured[2]. The Kane scheme[3], in which qubits are defined by nuclear spin states of buried phosphorus dopants in a silicon crystal, has also attracted considerable attention due to its promise of very long (ms or longer) decoherence times below 1K. Recent advances in single dopant fabrication[4-6], together with demonstration of fast single electron transistor (SET) charge detection[7,8], bring the Kane Si:P architecture closer to reality. These important results notwithstanding, the demonstration of single spin readout remains a major challenge. Here we consider a Si:P dopant-based qubit in which the logical information is encoded on the charge degrees of freedom. This system, which is complementary to the Kane concept, is not dependent on single-spin readout and, given the present availability of fabrication[4-6] and readout[7,8] technologies, can now be built. A two qubit gate based on the charge qubit scheme we describe will enable an experimental determination to be made of the key sources of decoherence and error in a nanoscale silicon QC architecture. Such devices therefore provide an important and necessary pathway towards the longer term goal of real-spin Si:P devices.

Semiconductor quantum dot charge-based qubits were first considered in 1995 by Barenco *et al.*[9], where quantum information was encoded in excitation levels, and later by Fedichkin *et al.*[10] for position-based charge qubits in GaAs. Very recently, coherent oscillations have been observed[11] in a GaAs double quantum dot providing realisation of a charge-based qubit with decoherence times above 1ns, accessible by existing fast pulse technology. In this paper we assess the potential of Si:P donor-based charge qubits by calculating the energetics and gate operation times for realistic device configurations and gate potentials and find that both one- and two-qubit operations times are well within the relevant decoherence times for the system.

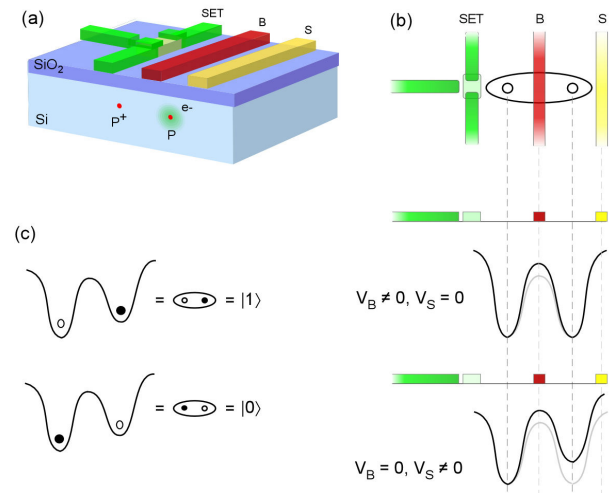

**Figure 1.** The buried charge qubit. (a) The solid-state charge qubit based on buried dopants D, forming a D-D$^+$ system with one electron, shown explicitly for the case for Si:P. (b) The gated charge qubit showing barrier (B-gate) and symmetry (S-gate) control, together with a single electron transistor (SET) for charge-based readout. (c) One possible choice of logical states $|0\rangle$ and $|1\rangle$ – defined as shown in terms of left and right-localised states.

The buried donor charge qubit is shown in Figure 1 for the case of P dopants in Si, although a number of other dopant-substrate systems could also be considered, such as GaAs:Si. The lowest two states of a single electron localised by the double well formed by two donor P$^+$ ions give rise to a natural identification of the quantum logic states. External control over the barrier height and potential off-set (or symmetry) is facilitated by B and S gates respectively, placed above the buried P-P$^+$ system, as in Figure 1(b). With appropriate negative bias we can identify localised qubit states with high precision: $|0\rangle = |L\rangle$ and $|1\rangle = |R\rangle$, as shown in Figure 1(c). Finally, a SET facilitates initialization and readout of the qubit.

The Si:P charge qubit will decohere faster than the Kane nuclear spin qubit – however, as the analysis of qubit dynamics will show, the typical gate operation times $\tau_{op}$ of order 50ps are also commensurately faster than the μs timescale[12] of the spin qubit. In what follows we estimate the decoherence time $\tau_\phi$ associated with phonons and gate fluctuations, finding $\tau_{op} << \tau_\phi$ for these mechanisms, and conclude that fluctuating background charges are more likely to dominate decoherence in most circumstances. Measurements on coupled GaAs quantum dots with 25-30 electrons per dot indicate $\tau_\phi > 1$ns[11,13]. Since such dots possess a similar vulnerability to background charge and may possibly couple more strongly to non-qubit space states than the one-electron Si:P system, we conclude that the coherence time for the buried charge qubit should be at least of order 1ns – certainly sufficient for proof-of-principle experiments on small-scale devices.

This paper is organised as follows. First, qubit dynamics are analysed to determine the fidelity of qubit states, and voltage pulses required for single-qubit operations. The processes for initialisation and SET readout are then outlined. Two possible qubit coupling schemes are described, and decoherence due to phonon mechanisms, gate fluctuations and charge traps is considered. Finally, fabrication of the charge qubit is described, and a possible scaled up N-qubit architecture is given.

The key to understanding single qubit gate operations is the effective Hamiltonian $H_Q$ describing the dynamics of the P-P$^+$ system in the presence of the S and B gates. In general, $H_Q$ will be of the form $H_Q = h_0(t) + h_x(t)\sigma_x + h_z(t)\sigma_z$, where the $\sigma_i$ operate in the basis of qubit states. The qubit logical states are defined by application of reference gate configuration voltages $(\bar{V}_B, \bar{V}_S)$ and are manipulated by fast-pulsed deviations $(\Delta V_B(t), \Delta V_S(t))$ from the reference configuration. Under these conditions, the time dependent coefficients can then be written as: $h_i(t) = C_0^{(i)} + C_S^{(i)}\Delta V_S(t) + C_B^{(i)}\Delta V_B(t)$, with $i=0, x, z$. The qubit dynamics are thus determined by the parameters $C_0^{(i)}$, $C_S^{(i)}$ and $C_B^{(i)}$ which depend explicitly on the donor separation $R$ and reference biases $\bar{V}_B$ and $\bar{V}_S$. For the device shown in Figure 1 the spatial dependence of the potentials induced in the silicon substrate due to the gate voltages was modelled using TCAD[14] for R = 27nm and these effective Hamiltonian parameters were computed by direct diagonalisation of the Hamiltonian in a basis of 12 molecular P-P$^+$ states with parameters appropriate to donor electrons in Si.

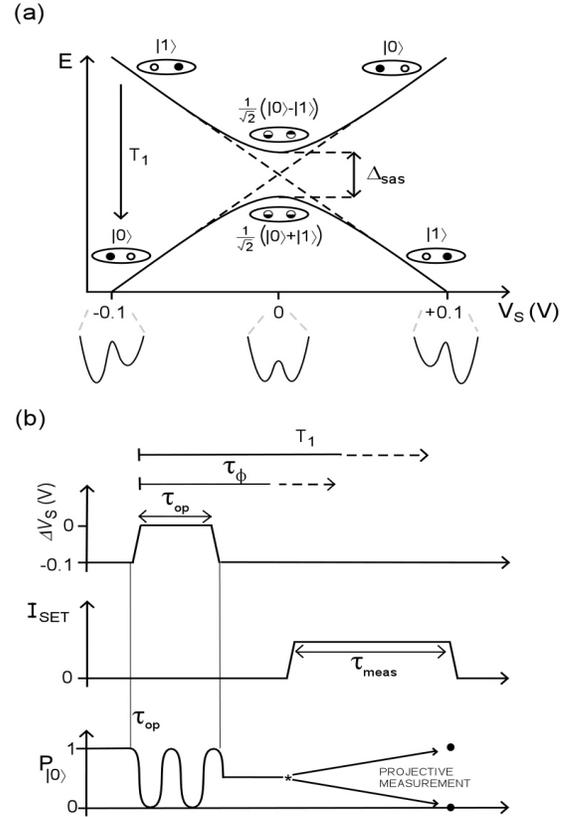

**Figure 2.** Qubit states and pulse timing. (a) Energy diagram illustrating the evolution of the eigenstates of the system as a function of applied S-gate bias. (b) Pulse timing diagram and SET readout showing the relative timescales for gate operation (τop), SET readout (τmeas), dephasing (τφ) and relaxation (T1).

We have two choices for the basis of logical qubit states corresponding to the lowest two states being localised or de-localised. Since SET readout is most easily carried out for localised states, we choose initially the configuration with non-zero S-gate bias, which defines our qubit states as $|0\rangle = |L\rangle$ and $|1\rangle = |R\rangle$. Careful examination of the lowest two eigenstates of $H_Q$ shows that for $\bar{V}_S \approx 0.1$V the qubit fidelity is optimal, with mixing of higher states less than $10^{-4}$. We discuss later the alternative delocalised basis choice $|0\rangle = |A\rangle$ and $|1\rangle = |S\rangle$, for which decoherence effects will be less severe. After setting the reference gate configuration to $(\bar{V}_B, \bar{V}_S) = (0$V, 0.1V), the gate bias pulses $(\Delta V_B(t), \Delta V_S(t))$ required for qubit control can be read off from $H_Q$. For example, a π/2 rotation over 50ps requires gate bias pulse values of ~(0.40V, +0.10V). The meaning of these values of $(\Delta V_B(t), \Delta V_S(t))$ is illustrated in the adiabatic state diagram of Figure 2(a): the double well potential is adjusted to the symmetric position $V_S = 0$, while at the same time raising the barrier to slow the Rabi oscillations down to the O(50ps) time scales accessible to state-of-the-art pulse generation.

Immediately after fabrication the qubit must be pre-initialised by removing one of the electrons from the P-P system to form the charge qubit. Using the S and B gates the electron in the right-hand donor well is ionised by a large S gate bias, at the same time the B-gate is raised to effectively isolate the electron in the left-hand well. After pre-initialisation, the SET conductance can be calibrated for the $|L\rangle$ and $|R\rangle$ states. Finally, initialisation of the charge qubit into the left state $|0\rangle$ is effected by simply biasing the S-gate and observing the SET conductance.

Prior to readout, the qubit is in a general state $|\psi\rangle = c_0|0\rangle + c_1|1\rangle$ resulting from a sequence of gate operations with the SET blockaded so that no current flows[15,16]. To perform a projective measurement a voltage is applied to the SET bias gate tuning it to a conductance peak – the current flow through the device decoheres the charge qubit strongly, and causes a transition in time $\tau_{meas} \ll T_1$ to a statistical mixture of the localised eigenstates (see Fig. 2b). Since the system has been calibrated in the pre-initialisation process, the SET will give a distinguishable reading[17] $I_{L,R}$ corresponding to the system having collapsed into the left or right state with probabilities $|c_0|^2$ and $|c_1|^2$ respectively.

In Figure 3 we present two distinct arrangements for qubit coupling, complete with gate structures and SET readout. For the case of the CNOT arrangement (Figure 3a) first proposed by Landauer[18] for quantum dot coupling, and by Fedichkin et al.[10] for GaAs qubits, the horizontal qubit $Q_1$ acts on the effective barrier height of the vertical qubit $Q_2$, and the coupling is primarily $\Gamma_{zx}\sigma_z^{(1)}\sigma_x^{(2)}$. We also consider here a CPHASE arrangement (Figure 3b) which is easily extended to a linear array of coupled qubits (Figure 4). Since the two qubits $Q_1$ and $Q_2$ act on each other symmetrically, quantum information can be transmitted either to the left or to right in a qubit array. The effective coupling for the CPHASE gate is $\Gamma_{zz}\sigma_z^{(1)}\sigma_z^{(2)}$. An in-depth investigation of the coupled qubit dynamics, controlled by such relatively complex gate structures, is beyond the scope of this paper, however, we have performed a preliminary semi-classical calculation to obtain an order of magnitude estimate. By moving a charge of 1.0 $e$ between the $a$ and $b$ positions of $Q_1$ for both coupling schemes (shown in Figure 3, with $Q_1$ chosen to be 30-60 nm from $Q_2$) the effective dynamics of $Q_2$ were determined with corresponding coupled qubit operation times of 0.1 → 1 ns.

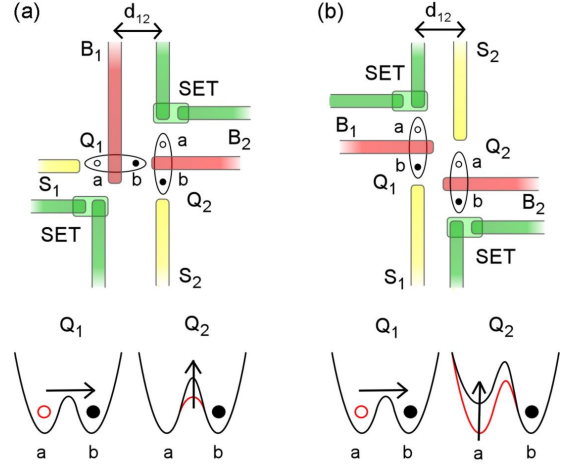

**Figure 3.** Qubit coupling schemes based on the Coulomb interaction. (a) CNOT configuration (b) CPHASE configuration.

Successful operation of quantum devices is contingent on coherence times remaining longer than the time required for arbitrary rotations. Primary sources of decoherence include phonons, Johnson noise on the gates, and materials-related charge noise. At mK temperatures the thermal phonon population is very small but spontaneous phonon emission can still occur. A calculation of LA phonon decoherence for the P-P$^+$ system[19] at 100mK concluded that for donor separations of 25nm and greater, $\tau^{phonon}$ is of order µs. The corresponding phonon-induced error rate for a one-qubit NOT gate (operating on $\sigma_x$) has recently been shown to be very low[20], while for $\sigma_x$ rotations a rate of 3x10$^{-3}$ was obtained[20]. This rate is, however, very sensitive to the phonon wavelength cut-off used in the calculation (in relation to the qubit length scale) and can be lower than 10$^{-5}$ for a wide range of parameters[21]. Irrespective of whether the error rate is below the 10$^{-4}$ level required for fault tolerant QC it appears that a significant number of gate operations will be possible, enabling proof-of-principle operation. An analysis of decoherence due to noise fluctuations on the S and B gates was carried out using a Master equation approach[22]. While the qubit is in a quiescent state, the dominant off-diagonal (od) contribution to the density matrix is $\rho_{od}(t) \approx \exp[-(C_S^{(z)})^2\lambda_S t/2\hbar^2]$, where $\lambda_S$ scales the fluctuations and is given by the Johnson formula: $\lambda_S = Rk_BT/\pi$. Using low temperature electronics at $T \sim 10$ K and $R \sim 50\Omega$, we obtain $\tau_\phi^{Johnson} \approx 2\pi\hbar^2/Rk_BTZ(C_S^{(z)})^2 \sim 720$ ns.

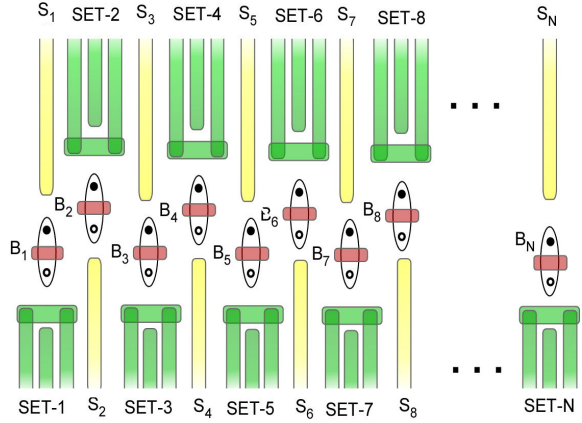

**Figure 4.** Schematic of a scaled-up architecture based on the staggered "CPHASE" configuration.

A serious source of decoherence for all charge-based qubits is due to charge fluctuations in the surrounding environment[23]. In particular, individual charge traps can produce sudden and large changes in the noise signal at random times (random telegraph signals). In superconducting charge qubits[2,24] the experimental coherence times of ~1ns are predicted to be limited by this charge noise[23], whilst the $\tau_\phi$ ~1ns observed in quantum dot qubits[11,13] may be similarly determined. We note that the nanosecond coherence times observed in GaAs qubits[11,13] are for quantum dots with ~25 electrons. The corresponding P-P$^+$ qubit coherence time might well be longer since it may be more difficult to isolate the qubit space from other states in GaAs quantum dots than for the simpler single electron qubit system. The use of high quality materials with low trap densities and refocusing pulse techniques may further extend the decoherence time. Furthermore, operation of the two-donor system in the delocalised basis where the qubit logic is less vulnerable to environmental charge fluctuations should lead to a significant suppression of the dephasing effects of charge noise, as is the case for Josephson 'phase' qubits where the coherence time has been extended to 500ns[25].

Realisation of the devices shown in Figures 1 and 3 requires an ability to dope a semiconductor at the single donor level with interdonor spacings in the range 20-100 nm. Due to the long range nature of the Coulomb coupling and the ability to tune the intra-qubit tunnelling rate using the B-gate, constraints on the donor spacings are significantly relaxed in comparison with previous spin-based donor schemes[3]. Single atom doping of a semiconductor with the required positional accuracy has recently been demonstrated by two contrasting approaches. In the first, scanning-probe lithography of a hydrogenated silicon surface together with epitaxial Si overgrowth are used to construct a buried P array with precision < 1 nm using atomic assembly[4,5]. In the second, the donors are implanted through an array of nanoscale apertures and on-chip ion impact detectors are used to ensure that just one P ion passes through each aperture[6]. The positional accuracy of the second approach is limited by the straggle which occurs during implantation and will be comparable to the donor depth (10-20 nm). Both approaches, although currently developed for Si:P, can in principle be applied to other systems, such as GaAs:Si. When combined with appropriate control and measurement electronics such devices allow gate voltage pulses on timescales < 50 ps and can perform single-shot projective measurements of electron position on timescales of order 10 ns-1μs[26,27]. It is therefore anticipated that one-qubit experiments on such structures will be possible soon.

With single atom doping and SET readout schemes for Si:P now available[6] it is expected that N-qubit architectures could be constructed in the near future. Figure 4 is a straightforward extension of the two-qubit CPHASE gate of Figure 3(b), where each qubit has an associated readout SET as well as the required S- and B-gates. The SETs would most conveniently be located on alternating sides of the one-dimensional array of qubits in order to localise the readout to the target qubit. Vertical via connections will be needed to make contact to the central B-gates, necessitating layered insulator and metal structures, which is standard in modern VLSI circuits.

In conclusion, gate operation times and decoherence rates have been calculated for Si:P charge qubits based on individual buried dopants with realistic gate configurations and bias voltages. Two coupling configurations were considered, including a CPHASE arrangement which can be conveniently scaled to a linear array of qubits. We find that one- and two-qubit gate operation times are accessible using existing pulse generator technology and are well within the estimates of decoherence due to phonons and gate fluctuations. The effect of environmental charge fluctuations can be gauged by measurements of $\tau_\phi$ ~ 1ns for GaAs quantum dot charge qubits[11,13], which can be considered as a lower bound for the Si:P qubit. While experimental measurement of decoherence times will be necessary to determine the viability of this scheme for fault-tolerent QC, it is clear that proof-of-principle demonstrations of qubit control and entanglement should be possible and as such will provide essential information towards the longer term goal of Si:P spin-based quantum computing.

This work was supported in part by the Australian Research Council, the Australian Government, the US National Security Agency, the Advanced Research and Development Activity and the US Army Research Office under contract number DAAD19-01-1-0653. We acknowledge helpful discussions with L. Fedichkin, H.S. Goan, C.I. Pakes and T.C. Ralph.